\documentclass{article}

\usepackage[nonatbib, final]{neurips_2021_ml4ps}
\usepackage{xcolor}

\usepackage[utf8]{inputenc} 
\usepackage[T1]{fontenc}    
\usepackage{hyperref}       
\hypersetup{
    colorlinks=true,
    linkcolor=cyan,
    filecolor=magenta,      
    urlcolor=blue,
    citecolor=blue,
}

\usepackage{url}            
\usepackage{booktabs}       
\usepackage{topcapt}        
\usepackage{amsfonts}       
\usepackage{nicefrac}       
\usepackage{microtype}      
\usepackage{lipsum}
\usepackage{graphicx}
\usepackage{amsmath}
\usepackage{multirow}
\usepackage{xspace}
\usepackage{cite}
\usepackage{listings}
\usepackage{xcolor}

\newcommand{\pt}{\ensuremath{p_{\mathrm{T}}}\xspace}
\newcommand{\kt}{\ensuremath{k_{\mathrm{T}}}\xspace}
\newcommand{\PX}{\ensuremath{\mathrm{X}}\xspace} 
\newcommand{\PY}{\ensuremath{\mathrm{Y}}\xspace} 
\newcommand{\Pq}{\ensuremath{\mathrm{q}}\xspace} 
\newcommand{\GeV}{\ensuremath{\,\text{Ge\hspace{-.08em}V}}\xspace}
\newcommand{\TeV}{\ensuremath{\,\text{Te\hspace{-.08em}V}}\xspace}
\newcommand{\PWpr}{\ensuremath{\mathrm{W}^{\prime}}\xspace} 

\begin{document}

\title{Particle Graph Autoencoders and Differentiable, Learned Energy Mover's Distance}

\author{Steven Tsan, Raghav Kansal, Anthony Aportela, Daniel Diaz,\\ 
\textbf{Javier Duarte, Sukanya Krishna, Farouk Mokhtar}\\
University of California San Diego\\
La Jolla, CA 92093, USA\\
\And
Jean-Roch Vlimant\\
California Institute of Technology\\
Pasadena, CA 91125, USA\\
\And
Maurizio Pierini\\
European Organization for Nuclear Research (CERN)\\
CH-1211 Geneva 23, Switzerland
}

\maketitle

\begin{abstract}
Autoencoders have useful applications in high energy physics in anomaly detection, particularly for jets---collimated showers of particles produced in collisions such as those at the CERN Large Hadron Collider. 
We explore the use of graph-based autoencoders, which operate on jets in their ``particle cloud'' representations and can leverage the interdependencies among the particles within a jet, for such tasks.
Additionally, we develop a differentiable approximation to the energy mover's distance via a graph neural network, which may subsequently be used as a reconstruction loss function for autoencoders.
\end{abstract}

\section{Introduction}

One of the primary motivations behind the development of the CERN Large Hadron Collider (LHC) was to search for new physics that may explain experimental observations left unaddressed by the standard model (SM) and expand our understanding of phenomena such as gravity and dark matter. 
The search for beyond the SM (BSM) physics has had limited success at the LHC possibly because current methods rely too heavily on hypothesized BSM signatures that may not reflect the true nature of the new physics.
To address this, there has been a growing interest in employing unsupervised machine learning (ML) models that can search for new physics independent of underlying signature assumptions.
For example, autoencoders, ML models that learn to map data down to a compressed encoding of its most salient features and then reverse such encodings back to their original form, have been employed for unsupervised anomaly detection~\cite{Monk:2018zsb,Cerri:2018anq,Farina:2018fyg,Cheng:2020dal,Kasieczka:2021xcg,Atkinson:2021nlt}. 
Autoencoders learn to accurately reconstruct data similar to what is seen during its training; however, anomalous signals rare or absent in the training data may not be accurately reconstructed---a property that can be used to detect them.

We propose particle graph autoencoders (PGAEs) based on graph neural networks (GNNs)~\cite{GNNsInHEP,Duarte:2020ngm} for unsupervised detection of new physics in multijet final states at the LHC. 
By embedding particle jet showers as a graph, GNNs are able to exploit particle-to-particle relationships to efficiently encode and reconstruct particle-level information within jets.
We posit that this can improve the capacity of autoencoders to learn a compressed representation of a jet and consequently help identify anomalous BSM multijet signal events from LHC data.
We also develop and validate a differentiable, learned approximation to an important distance metric, the energy mover's distance~\cite{emd}, using a GNN, dubbed EMD-GNN, which has the potential to be used as both a loss function to train a PGAE as well as a metric by which to judge how anomalous a jet is.

\section{Related Work}

\paragraph{Autoencoders in HEP}

A number of different autoencoder models have been studied for application in searching for new physics at the LHC~\cite{Monk:2018zsb,Cerri:2018anq,Farina:2018fyg,Cheng:2020dal}. 
One major drawback of many of these studies is the use of vector- or image-based representations of HEP data, which aren't well-suited to the sparsity and irregular geometry typical data produced at the LHC.
We propose instead to use the more natural set-based ``particle cloud''~\cite{Qu:2019gqs} representation for particles in a jet, which is inherently sparse and agnostic to the underlying geometry, and operate on this representation using GNNs. 

\paragraph{Graph networks}

GNNs are powerful, expressive networks that can operate on particle clouds and respect permutation invariance (for graph-level outputs) and covariance (for edge- and node-level outputs)~\cite{bronstein2021geometric}.
Due to this they have been steadily gaining prominence in HEP~\cite{GNNsInHEP}.
Notable examples include the dynamic graph convolutional neural network (DGCNN)~\cite{DGCNN}, which has been used for calorimetry in a high-granularity calorimeter~\cite{ExaTrkX} and jet identification~\cite{Qu:2019gqs}, as well as the interaction network~\cite{interactionnetwork} and its generalization to ``graph networks''~\cite{battaglia2018relational}, which have been used for jet identification~\cite{Moreno:2019bmu,Moreno:2019neq} and particle tracking~\cite{ExaTrkX,Dezoort:2021kfk}.
Other network architectures, GravNet and GarNet, have been studied for calorimetry~\cite{Qasim:2019otl,Iiyama:2020wap}.

GNNs for anomaly detection in HEP have not yet been fully explored.
However, recent work~\cite{Atkinson:2021nlt} develops an autoencoder-based strategy to facilitate anomaly detection for boosted jets, using a symmetric decoder that simultaneously reconstructs edge features and node features. 
Latent-space discriminators are used isolate W bosons, top quarks, and exotic hadronically-decaying exotic scalar bosons from QCD multijet background.
This work expands on that by performing a realistic resonance search using the PGAE model.

\subsection{Reconstruction loss functions}

Since the inputs and outputs are sets, a reconstruction loss needs to address the assignment problem, i.e. find a one-to-one correspondence between the two sets of nodes.
For a permutation-equivariant model, such as the presently considered GNN, the mean-squared error (MSE) is a standard choice because the order is preserved between the inputs and outputs.
The Chamfer loss~\cite{10.5555/1622943.1622971,Fan_2017_CVPR,Zhang2020FSPool} is permutation invariant, but has been found to be suboptimal~\cite{Kasieczka:2021xcg}.
Finally, the energy mover's distance (EMD)~\cite{emd}, related to the Earth mover's distance~\cite{emdimage,ot,POT}, is a desirable loss, which quantifies the difference between jets through optimal transport as the minimum ``work'' required to rearrange one jet into another by movements of transverse momentum between the particles in each jet.
Finding the EMD is a linear program~\cite{ot}, the exact solution to which is not efficiently differentiable, which limits its use directly as a loss function for training with backpropagation.
Thus, instead we develop a GNN-based approximation of the EMD, ``EMD-GNN'', which may be used in the future as a differentiable loss function.
Others have studied alternative approximations to the Earth or energy mover's distance to improve computability~\cite{Cai:2020vzx,fan2016point,sinkhorn}.


\section{Network Architectures}
\label{sec:architectures}

\paragraph{PGAE}
In the PGAE model, we represent input jets as fully-connected graphs where each constituent particle is represented as a node, and with edges between each pair of nodes.
When encoding and decoding, the graph structure of the data remains the same, but the node features, initially the particle's three-momentum $\boldsymbol{p} = (p_x, p_y, p_z)$, have their dimensionality reduced during the encoding phase.
We note the model can be expanded to consider additional particle-level information, such as particle type, electromagnetic charge, and pileup probability weight~\cite{Bertolini:2014bba}.
For the encoder and decoder, we use the edge convolution layer from Ref.~\cite{DGCNN}, which performs message passing along the edges and aggregation of messages at the nodes of the graphs.

The PGAE model is constructed using the PyTorch Geometric library~\cite{PyTorchGeometric}.
The node features inputted to the encoder are first processed by a batch normalization layer~\cite{batchnorm}.
The encoder itself is a single DGCNN layer, built from a fully connected neural network $\phi_\mathrm{e}$ with layers of sizes $(32, 32, 2)$ and rectified linear activation unit (ReLU) activation functions~\cite{relu}.
The network takes in as input $(\boldsymbol{p}_i, \boldsymbol{p}_j-\boldsymbol{p}_i)$, where $\boldsymbol{p}_i$ ($\boldsymbol{p}_j$) is the three-momentum for particle $i$ ($j$) and $i\neq j$.
The final layer produces a two-dimensional message vector from each pair of distinct particles.
These message vectors are aggregated (using a mean function) for each receiving particle using
\begin{equation}
\boldsymbol{h}_i = \frac{1}{|\mathcal N(i)|}\sum_{j\in \mathcal N(i)} \phi_\mathrm{e}(\boldsymbol{p}_i, \boldsymbol{p}_j-\boldsymbol{p}_i)\,,
\end{equation}
where $\mathcal N(i)$ is the neighborhood of particles connected to the $i$-th particle, which corresponds to all other particles in this case.
This summed message $\vec h_i$ is the bottleneck or encoded representation for the $i$-th particle.
The decoder is also a DGCNN layer, containing a network $\phi_\mathrm{d}$ with layers of sizes $(32, 32, 3)$ and ReLU activation functions after all but the final layer.
The input is a 3-dimensional vector representing $(\boldsymbol{h}_i, \boldsymbol{h}_j-\boldsymbol{h}_i)$ and the output is intended to reconstruct each particle's momentum.
We note that the architecture itself is covariant under permutations of the input particles and applicable to variable-size jets.

\paragraph{EMD-GNN}


The input to the EMD network is a pair of jets, represented in a single graph in a similar format to the PGAE's input, but with an extra binary channel per node to differentiate which jet it belongs to: $+1$ ($-1$) for the first (second) jet.

The EMD network itself is a GNN that utilizes three DGCNN layers, each one using two-layered fully connected networks with ReLU activations and batch normalization.
For each DGCNN layer the graph structure is dynamically recomputed with edges directed to each node from its 16-nearest-neighbors in feature space. 
A softplus activation is applied to the final output.

To ensure a symmetric distance metric, the network is inputted both permutations of the input jets and the predicted EMD value is the average of the network outputs.
We also utilize a symmetric loss function consisting of the MSE between the predicted and true EMD value, plus the MSE between the predicted EMDs for the two inputs.

\section{Experiments}
\label{sec:experiments}

\paragraph{Datasets}

The dataset~\cite{kasieczka_gregor_2019_3596919} comes from the LHC Olympics (LHCO) 2020 challenge and consists of a collection of simulated particle collisions divided up across three ``black boxes'' (BB) and one background QCD dijet events sample, each with one million particle collision events. 
Two of the black boxes (1 and 3) were injected with anomalous signals, while one (2) had no anomalous signals injected.
In addition, we also use a R\&D dataset from the LHCO~\cite{gregor_kasieczka_2019_2629073}, which has similar QCD events plus an additional 100,000 injected signal events with labels\footnote{Both datasets have been released under the CC-BY 4.0 license.}.

For input to the PGAE, we process the events using pyjet~\cite{Rodrigues:2019nct} to cluster $R=1$ anti-$\kt$ jets~\cite{Cacciari:2008gp, Cacciari:2011ma}, selecting only the leading two jets by transverse momentum per event, and then representing each jet as a vector of its constituents' three-momenta $p = (p_x, p_y, p_z)$, with array format $(N_\mathrm{particles}, 3)$.
We also require each jet to have $\pt>200\GeV$.
We train the autoencoders on the processed background dataset, and then evaluate them on the black boxes.
For the EMD-GNN, we use the same LHCO background data for training, but represent each particle in a jet by its relative $(\pt, \eta, \phi)$, forming all possible unique pairs of jets from 1,000 total events.
The true EMD value is computed with the EnergyFlow library~\cite{emd}, which bases its computation on the Python Optimal Transport library~\cite{POT}.
The dataset is randomly partitioned into training (80\%), validation (10\%), and testing samples (10\%).

\paragraph{PGAE+MSE Results}

\begin{figure}[htpb]
\centering
\includegraphics[trim={0 0 0 40pt}, width=0.47\textwidth]{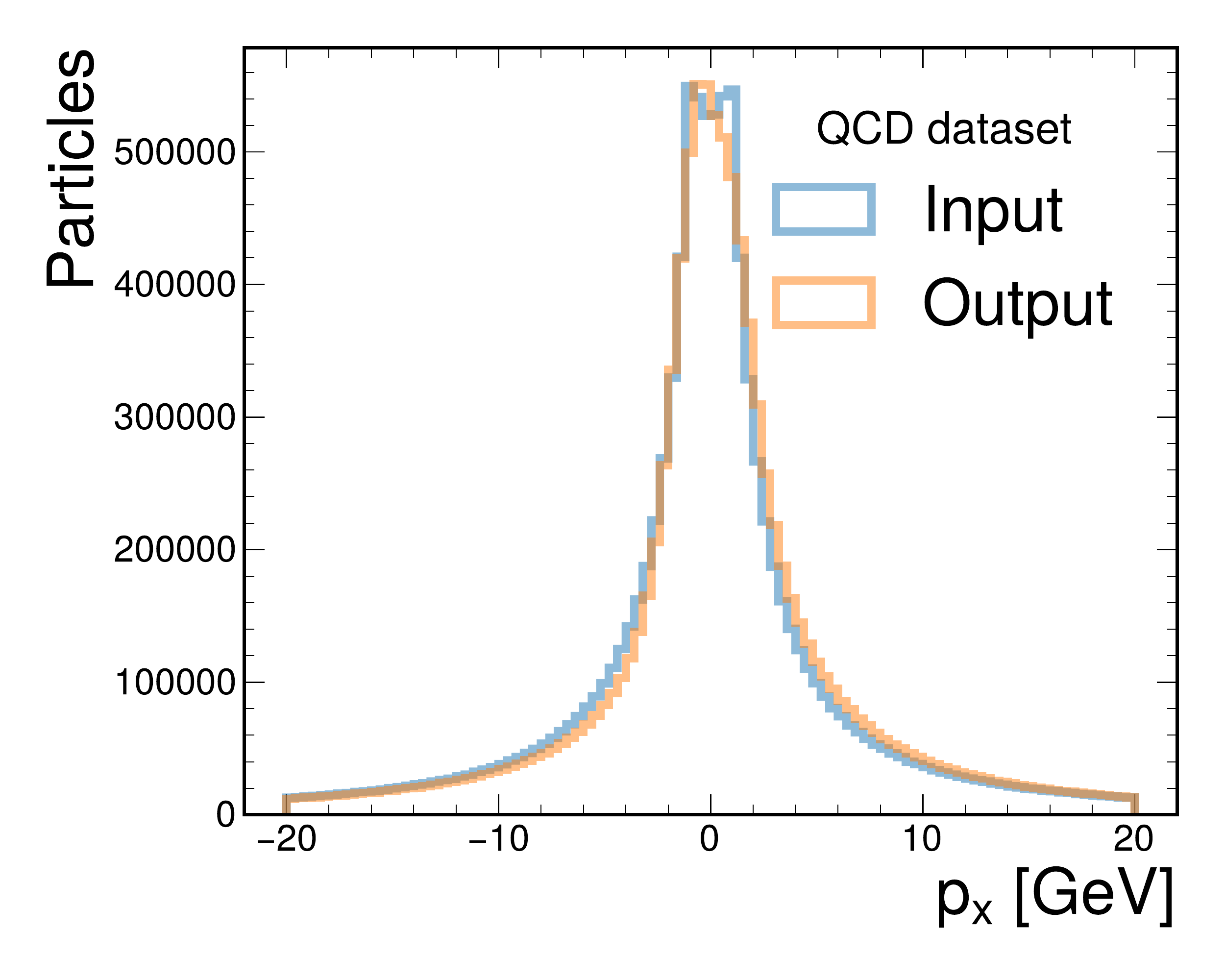}
\includegraphics[trim={0 0 0 50pt}, width=0.52\textwidth]{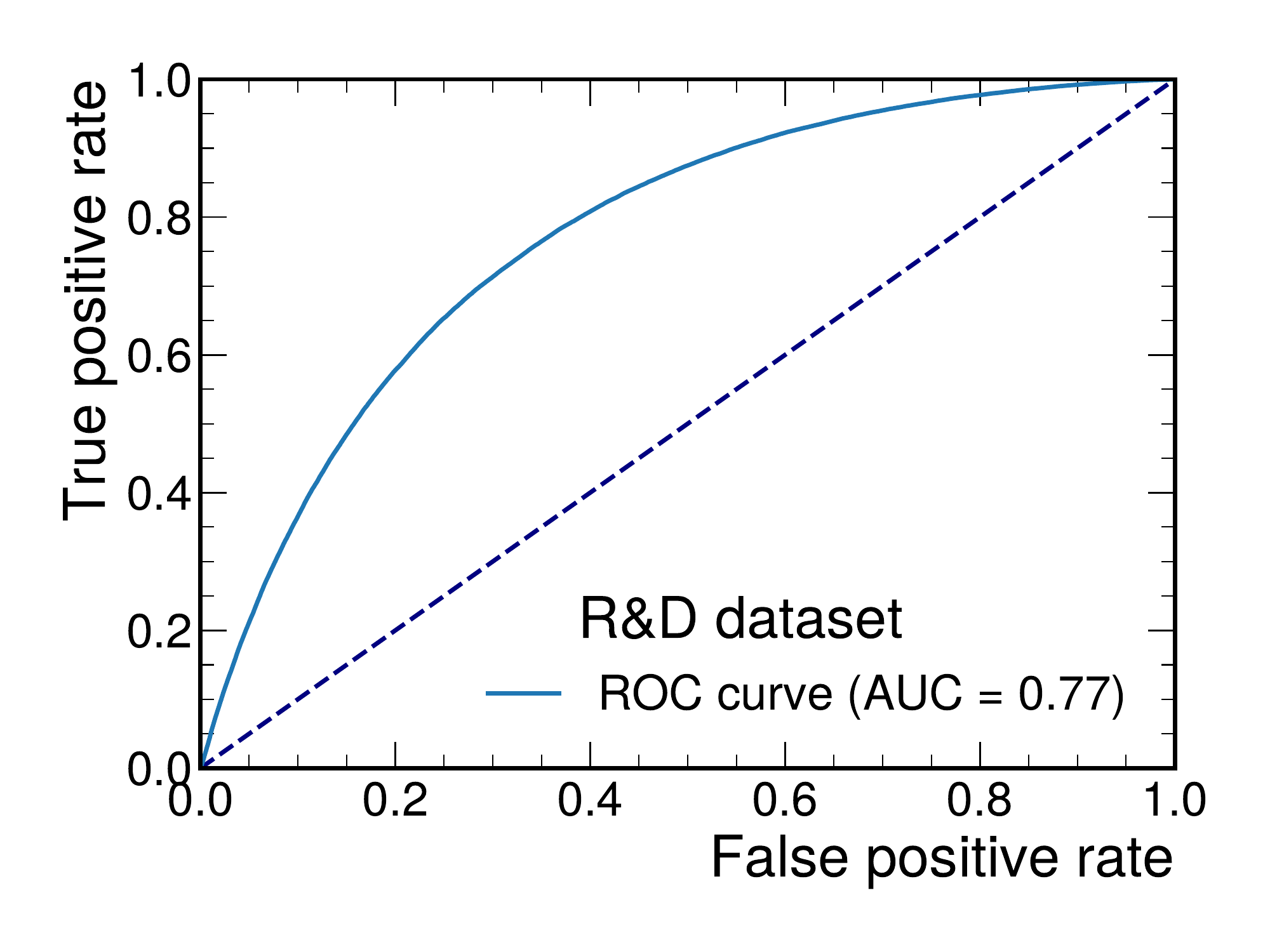}\\
\includegraphics[trim={0 0 0 30pt}, width=0.49\textwidth]{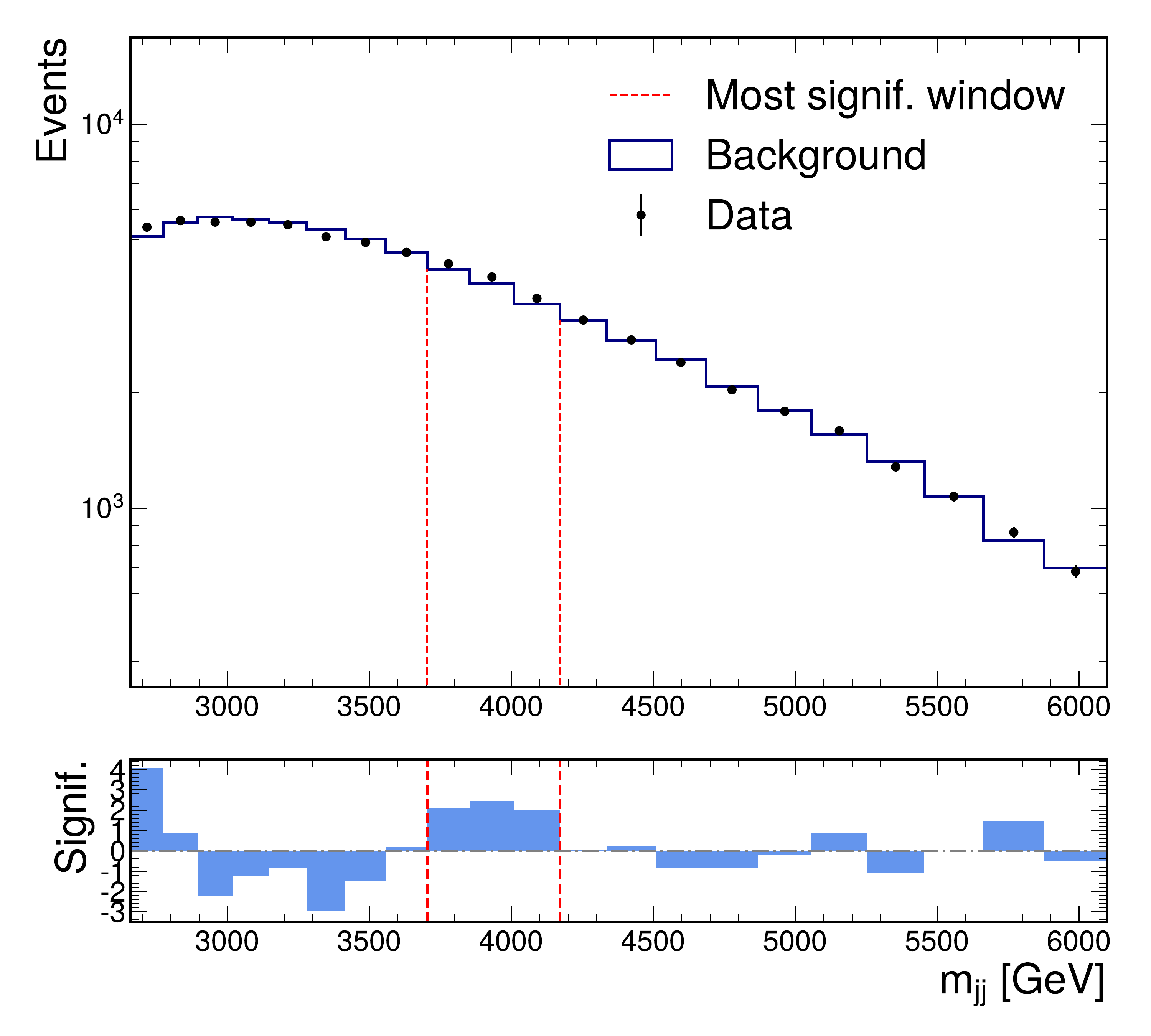}

\caption{Comparison of input and reconstructed distributions of a sample particle feature $p_x$ (top left) for the models trained with PGAE, evaluated on a test set.
ROC curve on the R\&D dataset for the PGAE model (top right), and the result of a resonance search using the dijet invariant mass performed on BB 1 (bottom).
For the search on BB 1, outlier jets have a reconstruction loss in the top 30\% and outlier events are required to have two outlier jets.
}
\label{fig:reco_anomdet}
\end{figure}

Training details for both the PGAE and EMD networks can be found in App.~\ref{app:training}.
Figure~\ref{fig:reco_anomdet} (left) shows a comparison of input and reconstructed features for the models trained with PGAE evaluated on a test set.
We see that the PGAE trained using a MSE loss performs well at reconstructing particles features. 
Although the model does not perfectly reconstruct the (double) peak in the center of the feature distributions, for the purpose of anomaly detection, this may not be a problem as long as non-outliers are reconstructed well enough that they have a lower reconstruction loss compared to actual outliers.


For anomaly detection we first study our algorithm on the R\&D dataset.
As the truth information is provided, we construct a receiver operating characteristic (ROC) curve to determine the effectiveness of the PGAE to identify a signal ($\PWpr\to\PX\PY$, $\PX\to\Pq\Pq$, and $\PY\to\Pq\Pq$ with $m_{\PWpr} = 3.5\TeV$, $m_\PX = 500\GeV$, and $m_\PY= 100\GeV$) that it did not observe during training. 
As seen in Fig.~\ref{fig:reco_anomdet} (center), the PGAE is capable of correctly identifying anomalies.

To evaluate the model's performance on unlabeled data, we perform a resonance search in the dijet invariant mass $m_\mathrm{jj}$, computed from the two leading jets per event, using a variable-width mass binning~\cite{Sirunyan:2018xlo} in the range from 2659\GeV to 6099\GeV.
We perform this dijet resonance search in BB 1, which contains a resonant dijet signal at $m_\mathrm{jj}\approx 3.8\TeV$.
We require both of the jets to be ``outliers,'' which we define as having a reconstruction loss exceeding a threshold corresponding to the 70\% quantile of the loss distribution on the evaluation dataset.
We note that because our algorithm is jet-focused, it is straightforward to generalize this search to multijet events.
To predict the background in the signal-enriched outlier region, we use the shape of the data in the background-enriched nonoutlier region.
We perform a maximum-likelihood fit to the ratio of the nonoutlier-to-outlier dijet mass distribution with a fourth-order polynomial to derive a transfer factor (TF) and take the nonoutlier data distribution weighted by the TF as an estimate of the expected background in the outlier region.
To derive the observed significance with the simplified background prediction, we use the bump hunter (BH) algorithm~\cite{Choudalakis:2011qn,pybumphunter} to look for resonances in windows spanning two to five bins.
With the MSE model in BB 1, we identify a possible resonance around $3.7\TeV$ with a global significance of $2.8\,\sigma$, which is consistent with the injected dijet resonance mass.

\paragraph{Comparison to other LHCO Contributions}

Many algorithms proposed for the 2020 LHCO~\cite{Kasieczka:2021xcg} performed similar evaluations on the R\&D and BB 1 dataset. 
Bump hunting in latent space ~\cite{bortolato2021bump} implemented a VAE that incorporates an invariant mass embedding in its latent space. 
This method obtained an AUC of 0.915 for an event-level discriminant on the R\&D dataset. 
Another algorithm combined a generative adversarial network (GAN) based autoencoder (AE) with the BH algorithm and achieved an AUC of about 0.90 on the R\&D dataset.
They also performed a dijet resonance search on BB 1, estimating a signal in the 3000--3600\GeV range. 
Another contribution used regularized likelihoods with manifold-learning flows~\cite{flows} to construct an anomaly score, which achieved an AUC of 0.7882 for their best performing model.
Tag N' Train ~\cite{amram2021tag} achieved an AUC of 0.918 on the full R\&D dataset, and detected a dijet resonance in BB 1 around 3800\GeV with a local significance of $4\,\sigma$. 
The deep ensemble anomaly detection method obtained an AUC of 0.96 using boosted decision trees, though we note that they performed semisupervised training on the R\&D dataset compared to our fully unsupervised training.

In comparison to our application of the PGAE, many of the other LHCO proposals achieved a higher AUC on the R\&D dataset.
However, because our discriminant is per jet, the discriminant values from multiple jets may be combined to achieve a better event-level discrimination. 
Moreover, many of the analysis methodologies are independent of the anomaly detection algorithm itself, thus are complementary to and can be integrated with our PGAE approach.

\paragraph{EMD-GNN Performance}

\begin{figure}[t]
\centering
\includegraphics[trim={0 0 0 45pt}, width=0.49\textwidth]{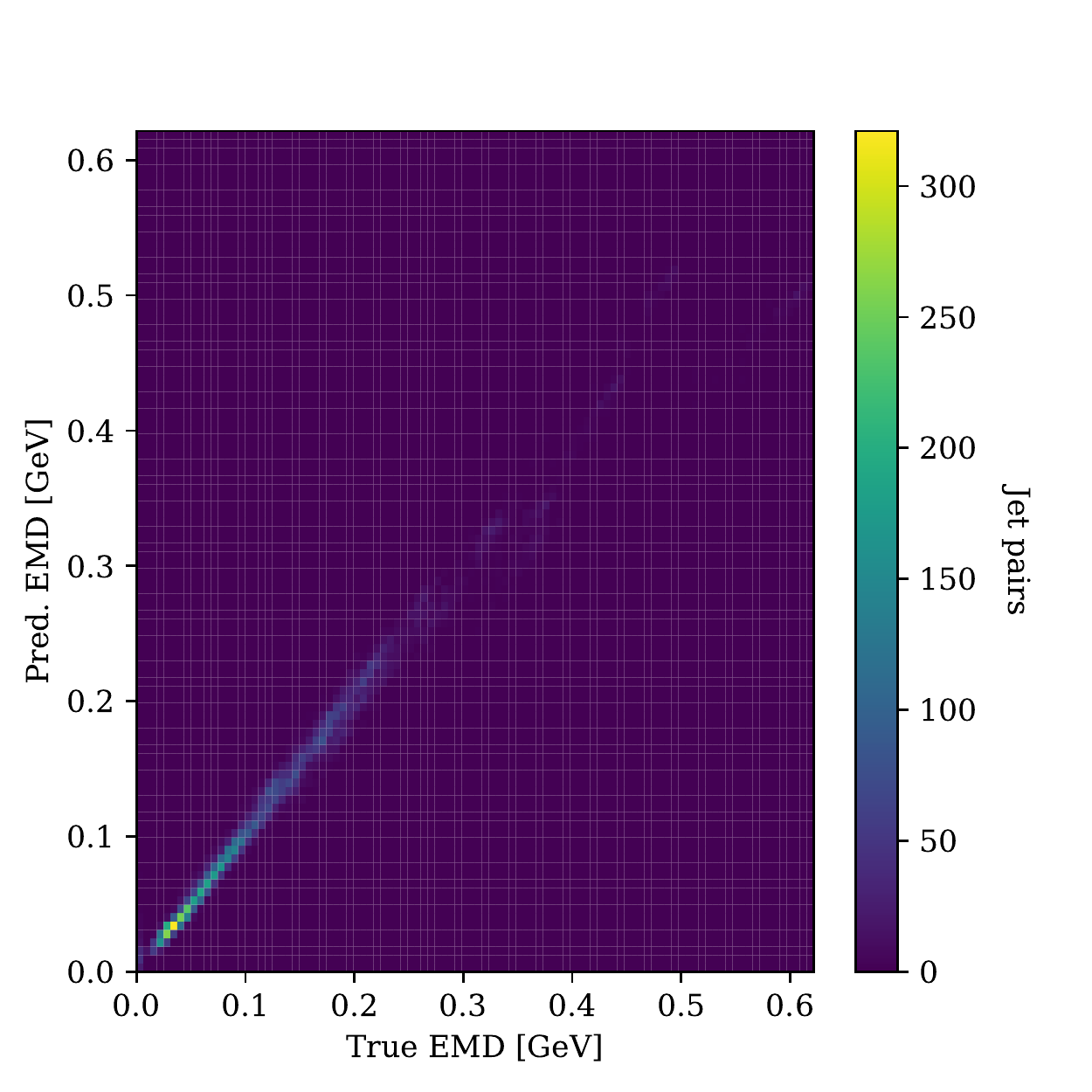}
\includegraphics[trim={0 0 0 45pt}, width=0.49\textwidth]{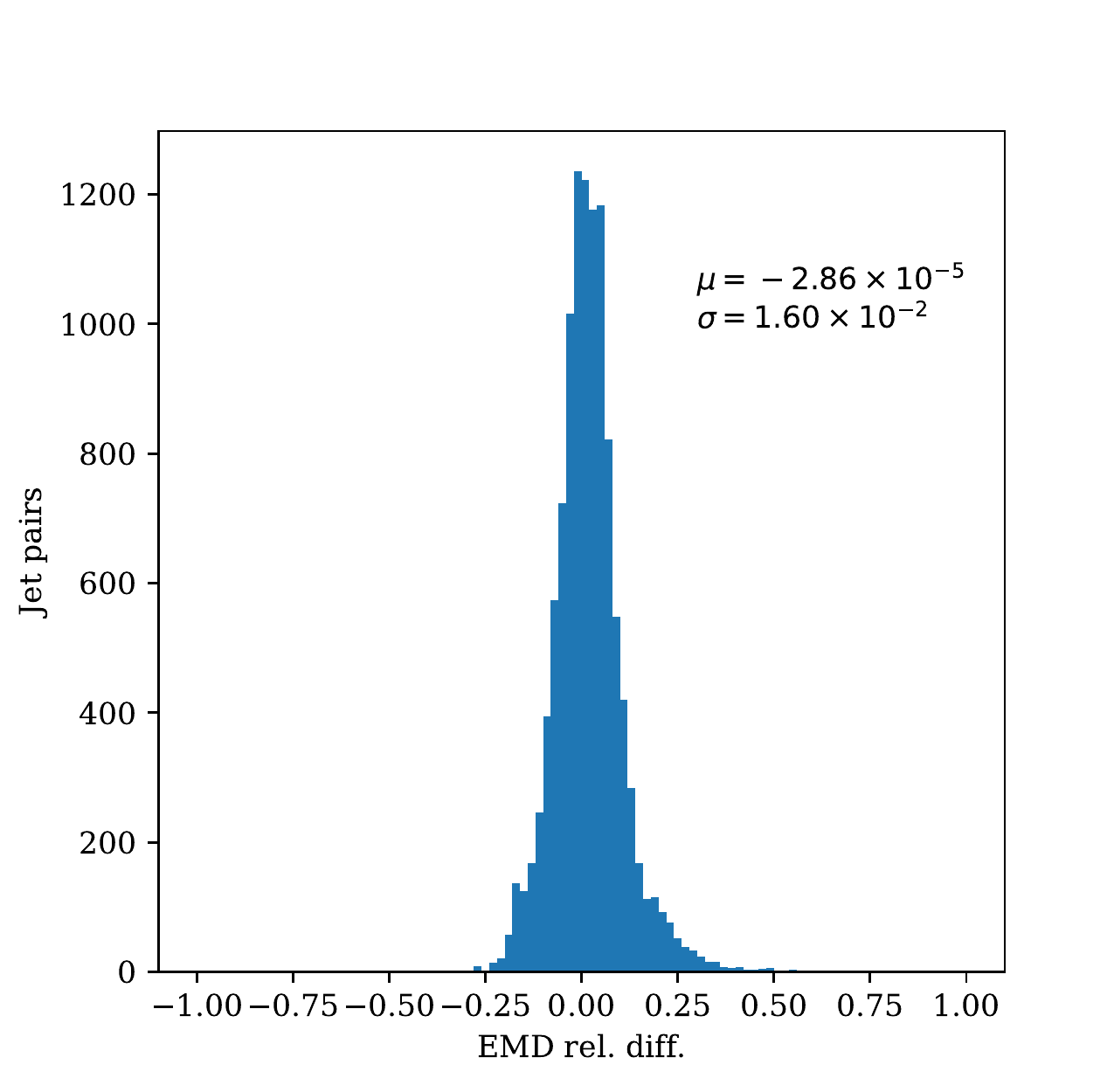}
\caption{Correlation between EMD-GNN's EMD prediction and the true EMD (left) and relative difference between corresponding true and predicted EMD (right) for the testing dataset composed of random pairs of jets.}
\label{fig:emdgnnperf}
\end{figure}

As shown in Fig.~\ref{fig:emdgnnperf}, the EMD-GNN can learn to approximate the EMD between real jets with a very high degree of accuracy, with a $-0.003\%$ relative difference between the predicted and true EMD on average, and a standard deviation of 1.6\%.
This indicates the potential to use this architecture to define a differentiable loss function for particle graph reconstruction.

\section{Summary}
We demonstrated that particle graph autoencoders (PGAEs) are effective at reconstruction of QCD background jets and, by extension, anomaly detection of anomalous jet signals. 
Good discrimination between background and signal jets was observed on the LHC Olympics (LHCO) R\&D dataset, which provides labels.
Moreover, using this algorithm, a dijet resonance was identified in the correct mass range in the LHCO Black Box 1 dataset with a global significance of $2.8\,\sigma$.
Additionally, we show that a graph neural network (GNN) can be used to approximate the energy mover's distance (EMD) and therefore potentially be used as a differentiable, permutation-invariant loss function. 
Future work will investigate optimizing the PGAE with the EMD-GNN as its loss function.

\paragraph{Broader Impact}
Our PGAE demonstrates the potential of unsupervised anomaly detection through particle graph representations of jets. 
As jet representations shift towards this particle cloud based format, permutation invariant loss functions such as the EMD become increasingly important. 
However approximations of EMD are known for their extremely large time complexity, as such, our EMD-GNN serves as an effective and differentiable alternative for approximating EMD.
In general, GNNs are becoming increasingly prevalent in many fields, including areas in which they may have harmful impacts on human welfare, to which this work may potentially contribute.

\begin{ack}
We thank the University of California San Diego Triton Research and Experiential
Learning Scholars (TRELS) program for supporting this research in its initial stages.
R.~K., A.~A., D.~D., J.~D., and F.~M. are supported by the US Department of Energy (DOE), Office of Science, Office of High Energy Physics Early Career Research program under Award No. DE-SC0021187 and by the DOE, Office of Advanced Scientific Computing Research under Award No. DE-SC0021396 (FAIR4HEP).
F.~M. is also supported by an Hal{\i}c{\i}o\u{g}lu Data Science Institute (HDSI) fellowship.
R.~K. is also supported by the LHC Physics Center at Fermi National Accelerator Laboratory, managed and operated by Fermi Research Alliance, LLC under Contract No. DE-AC02-07CH11359 with the U.S. Department of Energy (DOE).
J-R.~V. is supported by the DOE, Office of Science, Office of High Energy Physics under Award No. DE-SC0011925, DE-SC0019227, and DE-AC02-07CH11359.
M.~P. is supported by the European Research Council (ERC) under the European Union's Horizon 2020 research and innovation program (Grant Agreement No. 772369).
J-R.~V. is additionally supported by the same ERC grant as M.~P.
This work was performed using the Pacific Research Platform Nautilus HyperCluster supported by NSF awards CNS-1730158, ACI-1540112, ACI-1541349, OAC-1826967, the University of California Office of the President, and the University of California San Diego's California Institute for Telecommunications and Information Technology/Qualcomm Institute. 
Thanks to CENIC for the 100\,Gpbs networks.
\end{ack}

\clearpage

\appendix
\section{PGAE+MSE and EMD-GNN Training}
\label{app:training}

For training the PGAE, we use a batch size of 256, early stopping with a patience of 10 epochs, and an initial learning rate of 0.01.
Additionally we employ a learning rate scheduler that lowers the learning rate by a factor of 0.1 after every 4 epochs where the validation loss does not improve, with the minimum learning rate set by this process being $10^{-6}$. 
We use MSE as the loss in our experiments with the PGAE.

To train the EMD-GNN, we use the same training hyperparameters as the PGAE except for the batch size, which we set to 128.
Both models were trained on Nvidia GTX 1080Ti GPUs. 
The PGAE model requires about 5 days to train, and the EMD-GNN model takes about 20 hours.

\bibliographystyle{lucas_unsrt}
\bibliography{bibliography}

\end{document}